\documentclass[10pt,letterpaper]{article}
\usepackage{opex3}

\usepackage{amsmath,amssymb}

\DeclareMathOperator{\tr}{Tr}
\newcommand{\rf}{\text{rf}}
\newcommand{\dd}{\text{d}}
\newcommand{\V}[1]{{\boldsymbol{{\mathbf{#1}}}}}
\newcommand{\uv}[1]{\V{\hat{#1}}}

\newcommand{\ket}[1]{{\lvert#1\rangle}}

\begin{document}

\title{Nonlinear magneto-optical rotation in the presence of radio-frequency field}

\author{T. Zigdon,$^{1}$ A. D. Wilson-Gordon,$^{1}$ S. Guttikonda,$^{2}$ E. J. Bahr,$^{2}$ O. Neitzke,$^{3}$
S. M. Rochester,$^{3}$ and D. Budker$^{3,*}$ }

\address{$^{1}$Department of Chemistry, Bar-Ilan University,
Ramat Gan 52900, Israel\\ $^{2}$Department of Physics,
California State University East Bay, Hayward, California
94542-3084, USA
\\ $^{3}$Department of Physics, University of California at
Berkeley, Berkeley, California 94720-7300, USA}

\email{$^*$budker@berkeley.edu} 



\begin{abstract}
We report measurements of nonlinear magneto-optical rotation
(NMOR) for the D$_{2}$ line of $^{87}$Rb atoms in an
antirelaxation-coated vapor cell in the presence of a
radio-frequency (rf) field. The experimental NMOR signals as a
function of rf field frequency for various rf field powers are
compared to a theoretical model based on the density-matrix
formalism. The comparison between experiment and theory enables
understanding of the ground-state atomic spin polarization
dynamics, illustrated using plots of the probability
distribution of the atomic angular momentum.
\end{abstract}




\section{Introduction}

The dynamics of atomic spin polarization in the presence of
radio-frequency (rf) fields has been extensively studied (for
example, see \cite{KastlerJPR1950, CohenTannoudjibook1994,
WalkerRMP1997, Okunevich2004}). Recently, it has been
demonstrated that nonlinear magneto-optical rotation (NMOR) is
a sensitive probe of atomic spin dynamics \cite{KimballJAP2009,
BudkerPRA2000, BudkerRMP2002}, and can in fact be used to
selectively create and probe different multipole moments of the
atomic spin polarization \cite{YashchukPRL2003,
PustelnyPRA2006, AcostaOE2008}. There are numerous practical
applications for the use of NMOR to detect the response of
atomic spins to rf fields: for example, in nuclear magnetic
resonance (NMR) \cite{XuRSI2006}, including nuclear quadruple
resonance (NQR) \cite{GarrowayIEEE2001}, and magnetic resonance
imaging (MRI) \cite{XuPNAS0605396103} experiments. One can also
use such methods in tests of fundamental physics
\cite{BradleyRMP2003}. In our previous research, we built an
alkali-vapor magnetometer for the detection of rf fields
\cite{Ledbetter2007PRA}. Recent work by researchers using a
similar experimental setup is described in Refs.\
\cite{PhysRevLett.104.133601,PhysRevA.81.013422}. The
experiment \cite{Ledbetter2007PRA} demonstrated a sensitivity
to oscillating magnetic fields of 100 pG/$\sqrt{\textrm{Hz}}$.
The line shapes observed in the magnetometry experiment in the
limit of low rf power are well understood. However, when
higher-strength rf fields are applied, nontrivial line shapes
are seen, with similarities to those of the of the
``Majorana-Brossel effect'' in double-resonance spectroscopy
\cite{PhysRev.86.308}. We have conducted experimental and
theoretical investigations to explain the mechanism that
produces these line shapes.

In this work we present experimental NMOR signals taken on the
D$_{2}$ line of $^{87}$Rb atoms contained in an
antirelaxation-coated cell in the presence of an rf field. The
signals are primarily due to interaction with the
$F_{g}=2\rightarrow F_{e}=1$ transition (the subscripts $g$ and
$e$ indicate the ground and excited states, respectively). They
are compared to a theoretical model based on the density-matrix
formalism. The theoretical model, which is found to be in good
agreement with experimental results, enables the understanding
of the underlying atomic spin dynamics and their relationship
to the detected NMOR signals. The character of the observed
line shapes is found to be different depending on the value of
the Rabi frequency for the rf field, $\Omega_{\rf}$, relative
to the ground-state atomic-polarization relaxation rate
$\gamma_t$ and the magnetic resonance frequency $\Omega_L$.
This work is motivated by an ongoing project aimed at measuring
collisional transfer of alignment in collisions between
different ground-state alkali atoms
\cite{DFJacksonKimball2010}.

\section{Description of the experiment and theory}

\subsection{Description of experiment}
\label{subsec:experiment}

\begin{figure*}
\centering\includegraphics[width=0.8\textwidth]{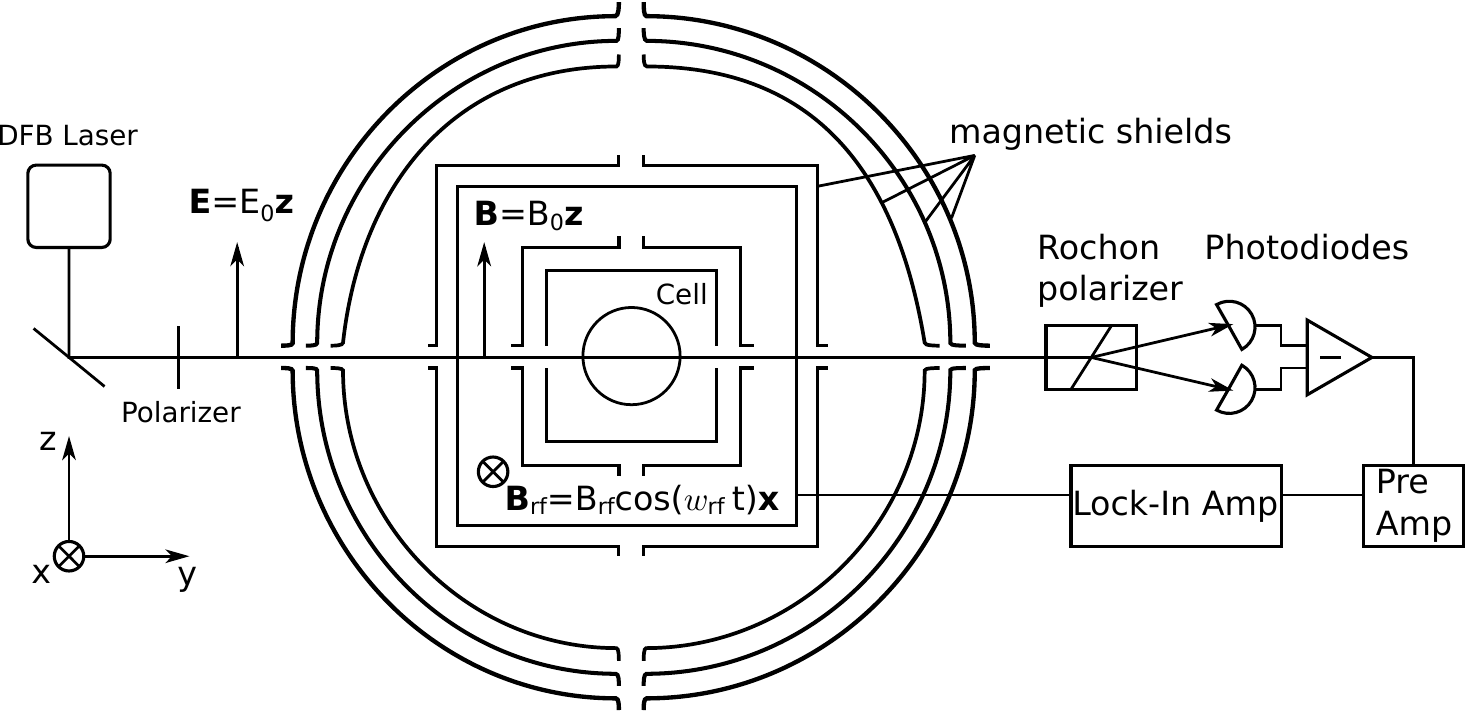}
\caption{A laser beam linearly polarized along the $z$ axis
propagates through a paraffin-coated vapor cell, producing
ground-state alignment in the $^{87}$Rb atoms. A constant
magnetic field $\V{B}_{0}$ and an oscillating rf magnetic field
$\V{B}_{\rf}$ are applied within the magnetic shield that
surrounds the cell. A balanced polarimeter and lock-in
amplifier are used to analyze polarization rotation of the
light induced by the atoms. A resonance is observed when the rf
frequency is equal to the ground-state Larmor frequency of the
atoms.}\label{fig:experimentalsetup}
\end{figure*}

The experiment employs a spherical paraffin-coated glass vapor
cell (diameter $=10$ cm) filled with a natural isotopic mixture
of rubidium. The cell coating allows polarization of
ground-state alkali atoms to survive several thousand wall
collisions \cite{PhysRev.147.41,PhysRevLett.81.5788}, thereby
extending the lifetime of atomic polarization. The cell is
heated by a constant air stream from a heat exchanger. During
all measurements in this work the temperature of the cell is
stabilized at $\approx41^\circ$C, resulting in a Rb vapor
density of $7.2(7) \times 10^{10}$ atoms per cm$^3$. The
density was extracted from fitting a low-light power ($\approx
8$ \textmu W) transmission spectrum for the Rb D$_2$ line to a
calculated spectrum assuming linear absorption. Under these
experimental conditions the longitudinal relaxation rate of
ground-state alignment of the isotope $ ^{87}$Rb was measured
to be $31.7(7)~\text{s}^{-1}$. The value represents the
effective relaxation rate of alignment polarization due to
diverse types of atomic collisions occurring in the vapor cell,
e.g.\ electron-randomization collisions with the wall, uniform
relaxation due to the reservoir effect and spin-exchange
collisions among Rb atoms \cite{PhysRevA.72.023401}. A
four-layer \textmu-metal magnetic shield (see Fig.\
\ref{fig:experimentalsetup}) surrounds the vapor cell and
reduces the external magnetic field by a factor of
$\sim\,$$10^{6}$ \cite{PhysRevLett.81.5788}. Additionally,
three orthogonal solenoidal coils placed inside the innermost
shield are used to compensate residual magnetic fields to below
$1$ \textmu G and to create additional static and oscillating
fields. For our measurements we apply a static magnetic field
$\V{B}_0$ in the $\uv{z}$ direction (typical strength $B_0=0.8$
mG) and an oscillating magnetic field
$\V{B}_{\rf}=B_{\rf}\uv{x}\cos \omega_{\rf}t$ along the
$x$-axis (typical amplitude up to $B_{\rf}=0.25$ mG) using the
sine wave output of a lock-in amplifier.

A laser beam, initially polarized along the $z$-axis,
propagates through the vapor cell in the $\uv{y}$ direction.
The beam is generated by a distributed feedback (DFB) laser,
which is tuned to the red wing of the $F_{g}=2\rightarrow
F_{e}=1$ transition in the $780$ nm D$_2$ line of $^{87}$Rb.
The wavelength is locked using a dichroic atomic vapor laser
lock \cite{Corwin1998AO, Yashchuk2000RSI}. The light intensity
is $10$ \textmu W throughout the measurements and the beam
diameter is $\approx 2$ mm. Linear dichroism of the atomic
medium induces changes in the light polarization. (For high
light power, the medium can also acquire circular birefringence
\cite{Budker2000PRL}.) After transmission through the vapor
cell the polarization of the light beam is analyzed using a
balanced polarimeter setup, consisting of a polarizing beam
splitter (Rochon crystal) and two photodiodes detecting the
intensities of the two beams exiting the crystal. The component
of the difference signal that oscillates at the frequency of
the rf field is then extracted by the lock-in amplifier.

\begin{figure}
\centering\includegraphics[width=5.2in]{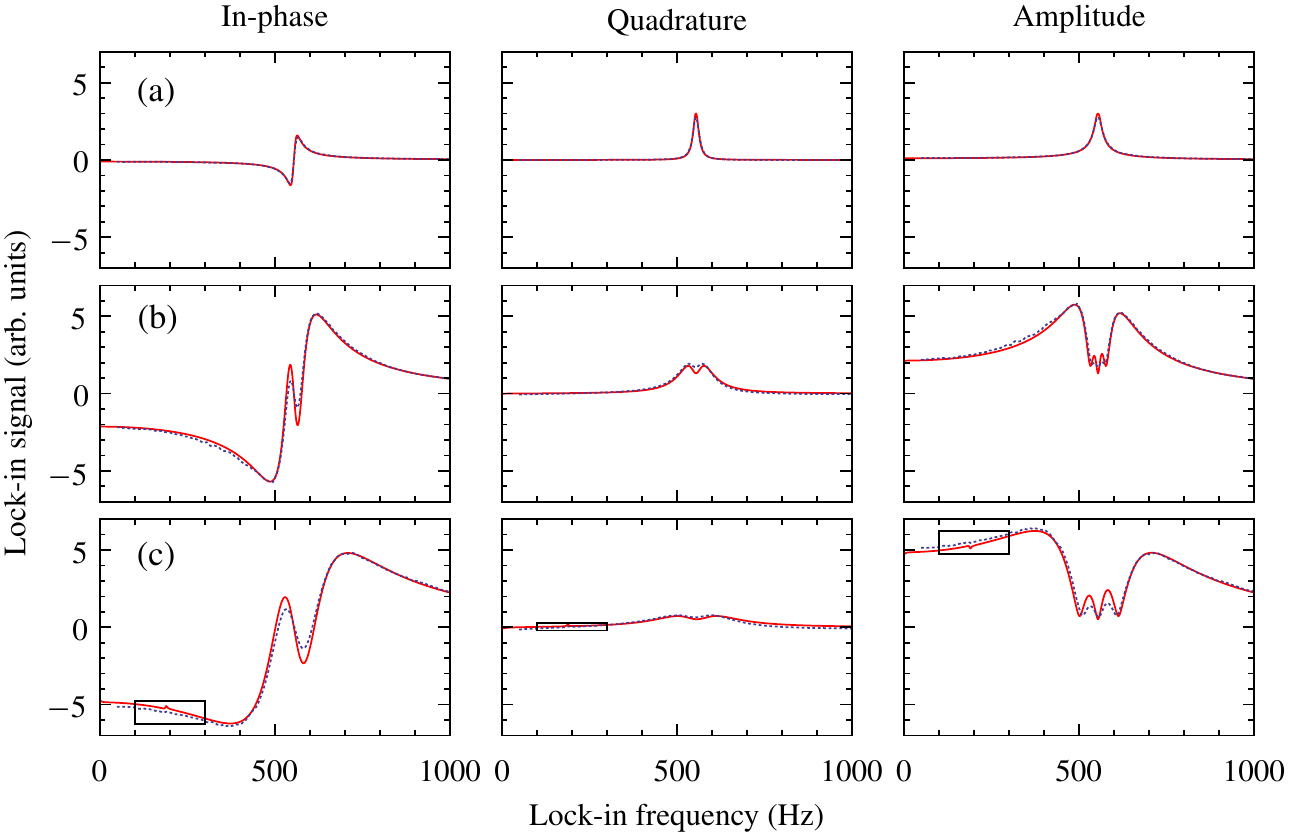}
\caption{Experimental spectra (dotted lines) and theoretical
predictions (solid lines) for three different values of the rf
field amplitude. The experimental values (obtained from
calibrated measurements of the coil current) in terms of the
Rabi frequency $\Omega_{\rf}$ are (a) $\Omega_{\rf}/(2\pi)=3$
Hz, (b) $\Omega_{\rf}/(2\pi)=63$ Hz, (c)
$\Omega_{\rf}/(2\pi)=158$ Hz. These values are used in the
theory, along with the parameters natural width
$\Gamma=2\pi\times6.1~\text{\textmu s}^{-1}$ (chosen equal to
the natural width of the Rb D$_2$ transition), light power $5$
\textmu W, ground-state relaxation rate
$\gamma_t=2\pi\times8~\text{s}^{-1}$, and bias-field Larmor
frequency $\Omega_L/(2\pi)=554.75$ Hz. These last three
parameters are chosen for optimal agreement with the
data---they are reasonably close to the experimental values.
The value for light power is applied to the theoretical model
using the method described in Sec.\ \ref{subsec:theory}. At low
rf-field strengths, as in row (a), the spectra are Lorentzians.
An additional central feature appears in the resonances when
$\Omega_{\rf}$ exceeds $\gamma_t$, as in rows (b) and (c).
Boxes in row (c) indicate regions plotted on expanded scales in
Fig.\ \ref{fig:HarmonicSpectra}.} \label{fig:Spectra}
\end{figure}

Experimental signals as a function of rf frequency, along with
predictions of the theory described in Sec.\
\ref{subsec:theory}, are given in Fig.\ \ref{fig:Spectra} for
different amplitudes of the rf field. At the lowest amplitudes,
the observed line shapes are Lorentzians, while for higher
amplitudes, additional features are seen.

\subsection{Description of theory}
\label{subsec:theory}

The experimental signal is primarily due to interaction with
the $F_g=2\to F_e=1$ transition of the D$_2$ line of $^{87}$Rb,
although there are also contributions from the $F_g=2\to F_e=2$
and $F_g=2\to F_e=3$ transitions. Theoretical modeling shows
that signals produced on each of these transitions have similar
line shapes, although the signal from a $F_g=2\to F_e=3$
transition is of the opposite sign. In fact, the effects that
we describe here are present for any transition with
ground-state angular momentum $F_g\ge1$. For the theoretical
treatment presented here, we therefore consider the simplest
case of a $F_{g}=1\to F_{e}=0$ transition, for which analytical
solutions are readily obtained. The model for this system gives
results in reasonable agreement with the experimental data,
indicating that the effects are not strongly dependent on the
angular momenta characterizing the transition.

The atoms are subject to a $\uv{z}$-directed field
$\V{B}_0=B_0\uv{z}$, corresponding to the Larmor frequency
$\Omega_{L}=g\mu_{B}B_{0}$, where $\mu_{B}$ is the Bohr
magneton and $g$ is the Land\'{e} factor (we set $\hbar=1$).
Linearly polarized light propagating in the $\uv{y}$ direction
with polarization in the $\uv{z}$ direction optically pumps the
system and creates an aligned state. An oscillating rf magnetic
field is applied in the $\uv{x}$ direction,
$\V{B}_{\rf}=B_{\rf}\uv{x}\cos\omega_{\rf}t$, corresponding to
the rf Rabi frequency $\Omega_{\rf}=g\mu_{B}B_{\rf}$. The
dependence on $\Omega_{\rf}$ of the rf line shape of the
optical rotation signal is studied. An analytic solution can be
obtained in case in which the rf power is low enough that the
rf power-broadened line width is much smaller than $\Omega_L$
and $\omega_{\rf}$. Numerical solutions are obtained in the
general case.

The experiment is performed using a vapor cell with an
antirelaxation coating. In such a cell, atoms can be optically
pumped in the light beam and then exit and return to the beam
after undergoing collisions with the cell walls, without the
polarization relaxing. Thus a complete theoretical description
must take into account the different conditions---and the
different state of the atoms---inside and outside the beam. If
the light power is low enough so that saturation effects do not
occur in the beam, however, the system can be modeled by
considering the average state of the atoms over the entire
cell. This amounts to mapping the case of a coated cell onto
that of an uncoated cell (i.e., one in which atomic
polarization can be considered to completely relax as soon as
the atoms leave the light beam). The light beam in the uncoated
cell is taken to have an intensity equal to the average
intensity over the entire cross-section of the coated cell, and
the atomic transit rate through the beam is taken to be equal
to the ground-state polarization relaxation rate in the coated
cell. (This ``effective'' uncoated cell must be very large, in
order to account for the slow ground-state relaxation of the
coated cell.) This is the case that we will consider.

Another complication arising in a complete model of an atomic
vapor cell is the velocity dependence of the atomic state due
to Doppler broadening and collisional velocity mixing effects.
These effects tend to change the dependence of a signal on the
light frequency. Since we hold the light frequency fixed in
this experiment, the main effect is the introduction of an
overall scaling factor. Therefore, we neglect the velocity
dependence in our model.

In the Zeeman basis \{$\ket{F_g=1,m=1}$, $\ket{F_g=1,m=0}$,
$\ket{F_g=1,m=-1}$, $\ket{F_e=0,m=0}$\}, the total
time-dependent Hamiltonian $H$ of the Doppler-free system under
the optical rotating-wave approximation is
\begin{equation}\label{TotalHamiltonian}
H = \left(
     \begin{array}{cccc}
       \Omega _L & \frac{\Omega_{\rf}}{2\sqrt{2}}(e^{i\omega_{\rf}t}+e^{-i\omega_{\rf}t}) & 0 & 0 \\
       \frac{\Omega_{\rf}}{2\sqrt{2}}(e^{i\omega_{\rf}t}+e^{-i\omega_{\rf}t}) & 0 & \frac{\Omega_{\rf}}{2\sqrt{2}}(e^{i\omega_{\rf}t}+e^{-i\omega_{\rf}t})  & -\frac{\Omega _R}{2\sqrt{3}} \\
       0 & \frac{\Omega_{\rf}}{2\sqrt{2}}(e^{i\omega_{\rf}t}+e^{-i\omega_{\rf}t}) & -\Omega _L & 0 \\
       0 & -\frac{\Omega _R}{2\sqrt{3}} & 0 & -\Delta \\
     \end{array}
   \right),
\end{equation}
where $\Omega _R$ is the Rabi frequency of the optical
transition induced by the linearly polarized light and
$\Delta=\omega-\omega_{0}$ is the optical detuning; $\omega$ is
the frequency of the light and $\omega_{0}$ is the frequency of
the ground to excited state transition in the absence of a
magnetic field. For the case in which the ground-state
relaxation rate and $\Omega_{\rf}$ are both much smaller than
$\Omega_{L}$ and $\omega_{\rf}$, we can also perform the
rotating-wave approximation on the rf field, in order to remove
the Larmor-frequency time dependence from the Hamiltonian. In
the rotating frame obtained using the unitary transformation
\begin{equation}\label{unitarytransformation}
U(t)=\left(
       \begin{array}{cccc}
         e^{-i\omega_{\rf}t} & 0 & 0 & 0 \\
         0 & 1 & 0 & 0 \\
         0& 0 & e^{i\omega_{\rf}t} & 0\\
         0 & 0 & 0 & 1 \\
       \end{array}
     \right),
 \end{equation}
the density-matrix evolution can be written in terms of an
effective Hamiltonian $H'=U^{-1}HU-iU^{-1}\frac{dU}{dt}$. After
dropping fast-oscillating, off-resonant terms, we have
\begin{equation}\label{undertransformationHamiltonian}
H'=\left(
  \begin{array}{cccc}
    -\Delta_{\rf} & \frac{\Omega_{\rf}}{2\sqrt{2}}& 0 & 0 \\
    \frac{\Omega_{\rf}}{2\sqrt{2}} & 0 & \frac{\Omega_{\rf}}{2\sqrt{2}} & -\frac{\Omega_{R}}{2\sqrt{3}} \\
    0 & \frac{\Omega_{\rf}}{2\sqrt{2}} & \Delta_{\rf} & 0 \\
    0 & -\frac{\Omega_{R}}{2\sqrt{3}} & 0 & -\Delta \\
  \end{array}
\right),
\end{equation}
where $\Delta_{\rf}=\omega_{\rf}-\Omega_{L}$ is the rf
detuning. The evolution of the density matrix $\rho$
(normalized so that $\tr\rho=1$) is described by the Liouville
equation
\begin{equation}\label{Liouvilleequation}
\dot{\rho}=-i[{H'},\rho]-\frac{1}{2}\{\zeta,\rho\}+\Lambda,
 \end{equation}
where [ ] denotes the commutator and $\{\: \}$ the
anticommutator.  The relaxation of the system is given by the
matrix
\begin{equation}\label{relaxation}
\zeta=\left(
       \begin{array}{cccc}
         \gamma_{t} & 0 & 0 & 0 \\
         0 & \gamma_{t} & 0 & 0 \\
         0& 0 & \gamma_{t} & 0\\
         0 & 0 & 0 &\Gamma+\gamma_{t} \\
       \end{array}
     \right),
 \end{equation}
where the excited state decays spontaneously with a rate
$\Gamma$ and the ground and excited states relax with a rate
$\gamma_{t}$ due to the exit of atoms from the light beam. The
matrix $\Lambda$ describes repopulation of the ground state due
to atoms entering the beam and spontaneous decay from the upper
state, and is given by
\begin{equation}
\label{repopulation} \Lambda=\left(
       \begin{array}{cccc}
         \frac{\gamma_{t}}{3}+\frac{\Gamma}{3}\rho_{e_{0}e_{0}} & 0 & 0 & 0 \\
         0 & \frac{\gamma_{t}}{3}+\frac{\Gamma}{3}\rho_{e_{0}e_{0}} & 0 & 0 \\
         0& 0 & \frac{\gamma_{t}}{3}+\frac{\Gamma}{3}\rho_{e_{0}e_{0}} & 0\\
         0 & 0 & 0 &0 \\
       \end{array}
     \right),
 \end{equation}
where the Zeeman ground and excited sublevels are denoted as
$g_{m_{g}}$ and $e_{m_{e}}$, and $\rho_{e_{0}e_{0}}$ is the
population in the excited state.

Under the rotating-wave approximation for the rf field
described above, the evolution equations for the rotating-frame
density matrix contain no explicit time dependence. We can
therefore set the time derivatives to zero and solve the
resulting system of linear equations for the steady state.
Taking the case of low light power, we solve the equations to
third order in the optical Rabi frequency, which is the lowest
order at which nonlinear optical rotation signals appear. Using
the inverse transformation $U^{-1}$, we transform back to the
laboratory frame to find the time-dependent density matrix.

The expectation value of the medium polarization is found from
the laboratory-frame density matrix. By substituting this value
into the wave equation, we can calculate the optical-rotation
signal measured in the laboratory frame in terms of the
rotating-frame density-matrix elements. After multiplying by
the reference signal and averaging over time, we find the
in-phase and quadrature (out-of-phase) signals per unit length
$\dd\ell$ of the medium to be
\begin{equation}\label{in-phasesignal}
\frac{\partial\varphi^{\text{in}}}{\partial\ell}
=-\frac{\sqrt{\frac{3}{2}}N\Gamma\lambda^{2}}{4\pi\Omega_{R}}\textrm{Im}(\rho_{e_{0}g_{-1}}-\rho_{e_{0}g_{1}}),
\end{equation}
\begin{equation}\label{quadraturesignal}
\frac{\partial\varphi^{\text{\text{out}}}}{\partial\ell}
=\frac{\sqrt{\frac{3}{2}}N\Gamma\lambda^{2}}{4\pi\Omega_{R}}\textrm{Re}(\rho_{e_{0}g_{-1}}+\rho_{e_{0}g_{1}}),
 \end{equation}
where $\lambda$ is the transition wavelength, $N$ is the atomic
density, and $\rho_{e_{0}g_{-1}}$ and $\rho_{e_{0}g_{1}}$ are
the optical coherences between the excited state and the ground
$g_{-1}$ and $g_{1}$ Zeeman sublevels, respectively. The
magnitude of the optical-rotation signal is given by
\begin{equation}\label{opticalrotation}
\begin{split}
\frac{\partial\varphi^{\text{abs}}}{\partial\ell}
&=\sqrt{\left(\frac{\partial\varphi^{\text{in}}}{\partial\ell}\right)^{2}+\left(\frac{\partial\varphi^{\text{out}}}{\partial\ell}\right)^{2}}\\
&=\frac{\sqrt{\frac{3}{2}}N\Gamma\lambda^{2}}{4\pi\Omega_{R}}
    \{[\textrm{Im}(\rho_{e_{0}g_{-1}}-\rho_{e_{0}g_{1}})]^{2}
+[\textrm{Re}(\rho_{e_{0}g_{-1}}+\rho_{e_{0}g_{1}})]^{2}\}^{1/2}.
\end{split}
\end{equation}

The expressions obtained after substituting in the solution for
the density matrix are complicated; to simplify the
presentation we assume that the light field is on resonance and
that $\Gamma$ is much greater than all other rates in the
problem. This gives
\begin{equation}\label{formulacoscoeff}
\frac{\partial\varphi^{\text{in}}}{\partial\ell}=
\frac{N\Delta_{\rf}\lambda^{2}\Omega_{\rf}(2{\gamma_{t}}^{2}+8\Delta_{\rf}^{2}-\Omega_{\rf}^{2})\Omega_{R}^{2}}{36\pi\Gamma\gamma_{t}
(\gamma_{t}^{2}+4\Delta_{\rf}^{2}+\Omega_{\rf}^{2})[4(\gamma_{t}^{2}+\Delta_{\rf}^{2})+\Omega_{\rf}^{2}]},
\end{equation}
\begin{equation}\label{formulasincoeff}
\frac{\partial\varphi^{\text{out}}}{\partial\ell}=
\frac{N\lambda^{2}\Omega_{\rf}(4\gamma_{t}^{2}+16\Delta_{\rf}^{2}+\Omega_{\rf}^{2})\Omega_{R}^{2}}{72\pi\Gamma
(\gamma_{t}^{2}+4\Delta_{\rf}^{2}+\Omega_{\rf}^{2})[4(\gamma_{t}^{2}+\Delta_{\rf}^{2})+\Omega_{\rf}^{2}]},
\end{equation}
where we have neglected the contribution to optical rotation
that is independent of the light power. Expanding these
expressions in a power series in $\Omega_{\rf}$, we obtain
\begin{equation}\label{expandedcoscoeff}
\frac{\partial\varphi^{\text{in}}}{\partial\ell}=
\frac{N\Delta_{\rf}\lambda^{2}\Omega_{R}^{2}}{72\pi\Gamma\gamma_{t}(\gamma_{t}^{2}+\Delta_{\rf}^{2})}\Omega_{\rf}
-\frac{N\Delta_{\rf}\lambda^{2}\Omega_{R}^{2}(7\gamma_{t}^{2}+10\Delta_{\rf}^{2})}{288\pi\Gamma\gamma_{t}(\gamma_{t}^{2}+
\Delta_{\rf}^{2})(\gamma_{t}^{2}+4\Delta_{\rf}^{2})}\Omega_{\rf}^{3}
+O[\Omega_{\rf}]^{5},
\end{equation}
\begin{equation}\label{expandedsincoeff}
\frac{\partial\varphi^{\text{out}}}{\partial\ell}=
\frac{N\lambda^{2}\Omega_{R}^{2}}{72\pi\Gamma(\gamma_{t}^{2}+\Delta_{\rf}^{2})}\Omega_{\rf}
-\frac{N\lambda^{2}
\Omega_{R}^{2}(4\gamma_{t}^{2}+7\Delta_{\rf}^{2})}{288\pi\Gamma(\gamma_{t}^{2}+\Delta_{\rf}^{2})
(\gamma_{t}^{2}+4\Delta_{\rf}^{2})}\Omega_{\rf}^{3}
+O[\Omega_{\rf}]^{5},
\end{equation}
\begin{equation}\label{expandedabscoeff}
\frac{\partial\varphi^{\text{abs}}}{\partial\ell}=
\frac{N \lambda ^2
\Omega _R^2 \Omega _{\rf}}{144 \pi  \Gamma  \gamma _t
\sqrt{\gamma _t^2+\Delta _{\rf}^2}}
-\frac{N \lambda ^2 \left(2
\gamma _t^2+5 \Delta _{\rf}^2\right) \Omega _R^2 \Omega
_{\rf}^3}{288 \pi  \Gamma  \gamma _t \left(\gamma _t^2+\Delta
_{\rf}^2\right)^{3/2} \left(\gamma _t^2+4 \Delta
_{\rf}^2\right)}
+O[\Omega _{\rf}]^5.
\end{equation}
These expressions describe resonances in $\Delta_{\rf}$
centered at $\Delta_{\rf}=0$. To lowest order in
$\Omega_{\rf}$, they are proportional to the real part,
imaginary part, and absolute value, respectively, of a complex
Lorentzian. Additional features appear at higher orders, as
discussed in the next section.

When $\Omega_{\rf}$ becomes of the same order as or exceeds
$\Omega_{L}$, the rotating-wave approximation for the rf field
is no longer valid. In this case we use the Hamiltonian $H$ of
Eq.\ \eqref{TotalHamiltonian} and proceed in the laboratory
frame. The Liouville equation is now time dependent; periodic
solutions can be found by expanding the density matrix in
Fourier series and retaining a finite number of harmonics. This
provides a linear system of time-independent equations that can
be solved numerically for the Fourier coefficients. The
observed optical rotation signals can then be found as before.

\section{Discussion}

The predictions of the density-matrix calculation described in
Sec.\ \ref{subsec:theory} are compared to the experimental data
in Fig.\ \ref{fig:Spectra}. For each value of the rf field
strength (characterized by the rf Rabi frequency
$\Omega_{\rf}$), the in-phase and quadrature components and the
magnitude of the optical-rotation signal as a function of the
rf frequency $\omega_{\rf}$ are shown. The version of the
theoretical treatment valid for arbitrary rf-field strength
discussed in Sec.\ \ref{subsec:theory} is used to generate the
theoretical predictions, although for the lowest rf power, the
signal is well described by the lowest-order terms of the
expansions \eqref{expandedcoscoeff}--\eqref{expandedabscoeff}.
Three regimes in the dependence on $\Omega_{\rf}$ can be
identified. At the lowest field strengths,
$\Omega_{\rf}<\gamma_t$, the in-phase and quadrature resonances
in rf frequency take the form of dispersive and absorptive
Lorentzians of characteristic width $\gamma_t$ (Fig.\
\ref{fig:Spectra}a). At intermediate field strengths,
$\gamma_t<\Omega_{\rf}<|\Omega_L|$ (we assume
$\gamma_t\ll|\Omega_L|$), the Lorentzians broaden and
additional narrow features are seen at the center of the
resonances (Fig.\ \ref{fig:Spectra}b), the result of
polarization-averaging effects discussed below.

\begin{figure}
\centering\includegraphics{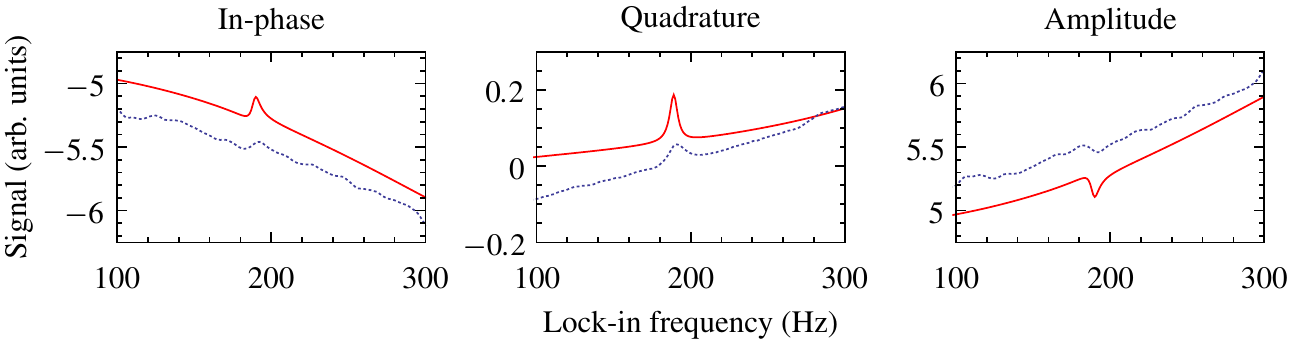}
\caption{Row (c) of Fig.\ \ref{fig:Spectra}
plotted on expanded scales. (The vertical scale is different
for each plot.) A resonance due to high-field effects can be
seen near
$\omega_{\rf}/(2\pi)=\frac{1}{3}\Omega_L/(2\pi)\approx185$ Hz.}
\label{fig:HarmonicSpectra}
\end{figure}

For higher fields, $\Omega_{\rf}>|\Omega_{L}|$, effects due to
ac Zeeman shifts and far-off-resonant fields are predicted to
become important. We did not perform measurements in this
regime, but the beginning of these effects can be seen in the
data (Fig.\ \ref{fig:Spectra}c). The negative-frequency
component of the rf field results in a resonance at
$\omega_{\rf}=-\Omega_L$ symmetric to the one at
$\omega_{\rf}=\Omega_L$. In Fig.\ \ref{fig:Spectra}(b) and (c),
the off-resonant tail of this negative-frequency resonance
produces an overall slope in the in-phase component of the
positive-frequency signal. Calculations for values of
$\omega_{\rf}$ of the same order as $\Omega_L$ predict
higher-order resonances at odd fractions ($1/3$, $1/5$, etc.)
of $\Omega_L$. When the experimental data presented in Fig.\
\ref{fig:Spectra}(c) are plotted on expanded scales, as in
Fig.\ \ref{fig:HarmonicSpectra}, a higher-order resonance can
be observed at one third the frequency of the main resonance,
in agreement with theoretical predictions.

\subsection{Low-field regime}

We first consider the low-rf-field regime,
$\Omega_{\rf}<\gamma_t$. This case was discussed in Ref.\
\cite{Ledbetter2007PRA}. An example of experimental data taken
in this regime compared to theoretical predictions is shown in
Fig.\ \ref{fig:Spectra}(a). As described by Eqs.\
\eqref{expandedcoscoeff}--\eqref{expandedabscoeff}, the
resonance observed in the in-phase component, quadrature
component, and magnitude of the optical-rotation signal takes
the form of the real (dispersive) part, imaginary (absorptive)
part, and the magnitude of a complex Lorentzian, respectively.
The absolute value of a complex Lorentzian has the line shape
of the square root of the absorptive part. Near resonance the
quadrature component is the primary contributor to the
magnitude, while farther from resonance, the in-phase component
provides the main contribution. The characteristic width of the
observed resonances is determined by the ground-state
relaxation rate.

\begin{figure}
\centering\includegraphics{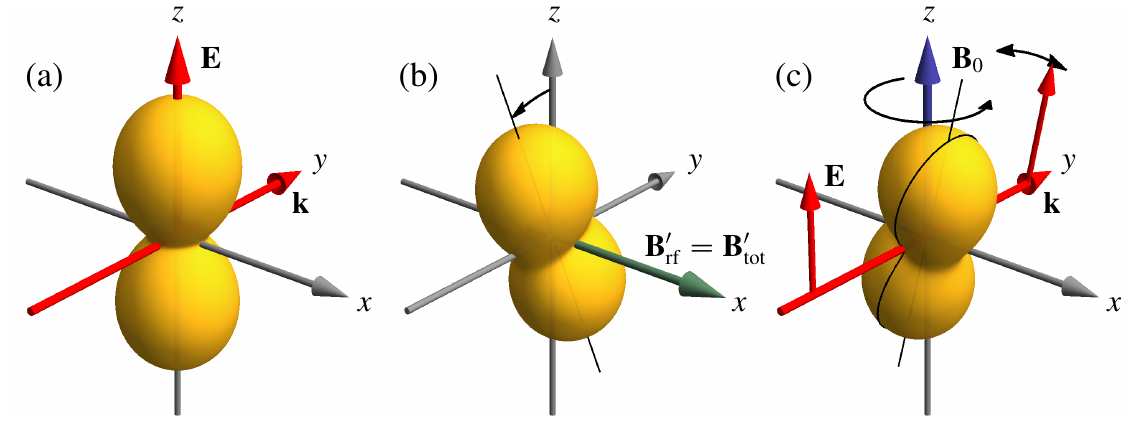}
\caption{Angular-momentum-probability surfaces
illustrating the behavior of the system with a resonant
($\Delta_{\rf}=0$), weak rf field. (a) The light, with electric
field $\V{E}$, produces atomic alignment along the $z$-axis.
(b) In the rotating frame, the atoms precess around the
$x$-axis. Due to relaxation, a steady state is reached. (c) In
the laboratory frame, the atoms precess around the $z$-axis,
with the linear dichroism of the ensemble inducing oscillating
optical rotation in the $z$-polarized, $\uv{y}$-propagating
light field. (Media 1 animates the precession and optical
rotation as a function of time.) The transmission axis of the
polarized ensemble, in this case parallel to the atomic
alignment axis, is indicated with a line drawn on the surface.}
\label{fig:LowPowerOnRes}
\end{figure}

Optical pumping by the light field removes atoms from the
$m_g=0$ sublevel, leaving an incoherent mixture of atoms in the
$m_g=\pm1$ sublevels. The atomic polarization can be
illustrated using the angular-momentum-probability surface
(AMPS) \cite{Auz97,Roc2001,ABRBook}, whose radius in a given
direction is determined by the probability of measuring the
maximum possible angular-momentum projection in that direction.
This provides the quantum-mechanical analog of the classical
angular-momentum probability distribution. The optically pumped
distribution corresponds to atomic alignment along the $z$-axis
with a ``peanut''-shaped probability distribution (Fig.\
\ref{fig:LowPowerOnRes}a). (All of the AMPS shown here are
obtained directly from the density-matrix calculation. A
quantity of the lowest-rank, isotropic polarization moment is
subtracted from each figure so that the anisotropic
polarization can be more clearly seen \cite{ABRBook}.) Because
the $m_g=0$ sublevel has been depleted, the atomic medium
transmits $z$-polarized light, while tending to absorb
orthogonally polarized light---i.e., the atoms function as a
polarizing filter with transmission axis along the atomic
alignment axis \cite{Kan93}. This linear dichroism can induce
rotation of the light polarization if the transmission axis is
tilted away from the light polarization axis.

The observed signals in the low- and intermediate-field regimes
can be most readily understood in terms of Larmor precession of
the atomic alignment in the combined static and rf magnetic
field. In the rotating frame, under the rotating-wave
approximation for the rf field, the effect of the magnetic
fields $\V{B}_0$ and $\V{B}_{\rf}$ can be described in terms of
fictitious static fields $\V{B}'_0$ and $\V{B}'_{\rf}$. These
fields can be determined by examining the rotating-frame
Hamiltonian \eqref{undertransformationHamiltonian}. The bias
field in the rotating frame, $\V{B}'_0$, points along
$\V{B}_0$, while its field strength is such that it produces a
Larmor frequency given by the detuning of the rf field from
resonance: $\Omega'_L=-\Delta_{\rf}=-\omega_{\rf}+\Omega_L$.
The rf field $\V{B}_{\rf}$ becomes a static field
$\V{B}'_{\rf}$ in the $xy$ plane with associated Larmor
frequency $\Omega'_{\rf}=\Omega_{\rf}/2$. The direction of this
field in the $xy$ plane depends on the arbitrary phase chosen
for the rotating frame; in our convention $\V{B}'_{\rf}$ points
along $\uv{x}$. Thus the resultant field
$\V{B}'_{\text{tot}}=\V{B}'_0+\V{B}'_{\rf}$ lies in the $xz$
plane.

The atomic polarization evolves in the rotating frame under the
action of $\V{B}'_{\text{tot}}$. For low field strengths, this
evolution is Larmor precession, so that the polarization
continues to correspond to alignment. Because all of the
external fields are static in the rotating frame, the effect of
relaxation leads to a steady state for the atomic polarization.
First consider the case in which the rf field is exactly on
resonance ($\Delta_{\rf}=0$). Then $\V{B}'_{\text{tot}}$ is
equal to $\V{B}'_{\rf}$ and points in the $\uv{x}$ direction
(Fig.\ \ref{fig:LowPowerOnRes}b), and the atomic alignment
precesses in the $yz$ plane. Under the assumption
$\Omega_{\rf}\ll\gamma_t$, the precession frequency is much
less than the relaxation rate, so that each atom precesses
through a small angle before relaxing. The rotating-frame
steady-state ensemble polarization thus consists of alignment
at a small angle to the $z$-axis in the $yz$ plane (Fig.\
\ref{fig:LowPowerOnRes}b).

In the laboratory frame, the alignment precesses about the
$z$-axis (Fig.\ \ref{fig:LowPowerOnRes}c). At the instant that
the alignment is in the $yz$ plane, it does not induce any
polarization rotation in the $\uv{y}$-propagating light field.
On the other hand, whenever the alignment axis is tilted away
from the initial light polarization axis (the $z$-axis) in the
plane transverse to the light propagation direction (the $xz$
plane), the atoms can induce optical rotation. Because of the
precession of the alignment, the optical-rotation signal
oscillates. The amplitude of the signal is determined by the
amount of the alignment and the angle between the alignment
axis and the $z$-axis (zenith angle). The phase of the
oscillating signal is determined by the angle of the alignment
axis about the $z$-axis in the rotating frame (azimuthal
angle). The direction of the alignment axis in the rotating
frame corresponds to its direction in the laboratory frame when
the rf field is maximum. Only the component in the $xz$ plane
will induce optical rotation in $y$-propagating light, so an
alignment axis in the $xz$ plane in the rotating frame produces
a signal in phase with the rf field oscillation, while the
component in the $yz$ plane produces a quadrature component in
the signal. In the case of Fig.\ \ref{fig:LowPowerOnRes} the
optical-rotation signal is entirely in the quadrature
component.

\begin{figure}
\centering\includegraphics{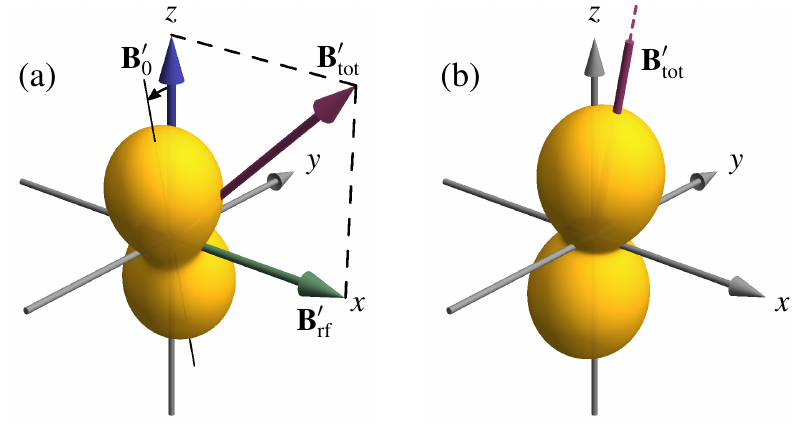}
\caption{As Fig.\ \ref{fig:LowPowerOnRes}, but
with an off-resonant rf field. (a) In the rotating frame, the
atoms precess around the total effective field, which lies in
the $xz$ plane. As a result, the polarization is no longer
entirely in the $yz$ plane. (b) Far off resonance, the
effective magnetic field is large enough that the polarization
is completely averaged about the magnetic-field direction. (The
arrow representing $\V{B}'_{\text{tot}}$ is truncated, as it is
too long to fit entirely on the plot.) Media 2 shows the
polarization, along with the in-phase component of the
optical-rotation signal, as the rf-field detuning is swept
through resonant and off-resonant conditions} \label{fig:LowPowerOffRes}
\end{figure}

As the rf field is tuned away from resonance, the total field
in the rotating frame begins to point away from the $x$-axis
and toward the $z$-axis. Precession about this field then takes
the alignment in the rotating frame out of the $yz$ plane
(Fig.\ \ref{fig:LowPowerOffRes}a).
This tends to reduce the angle that the steady-state alignment
makes with the $z$-axis, reducing the amplitude of the
laboratory-frame optical rotation signal. On the other hand,
because the alignment now has a component in the $xy$ plane in
the rotating frame, the oscillating rotation signal gains an
in-phase component. When the rf field is tuned far enough away
from resonance so that $|\Delta_{\rf}|>\gamma_t$, the
precession frequency in the rotating frame becomes large enough
that the atoms undergo an entire precession cycle before
relaxing. The ensemble polarization is then averaged about the
direction of the total magnetic field (Fig.\
\ref{fig:LowPowerOffRes}b). Because $\V{B}'_{\text{tot}}$ lies
in the $xz$ plane, the signal is now predominately in phase
with the rf field, and the quadrature component is strongly
suppressed. As the detuning becomes large, the average
polarization points more and more along the $z$-axis, and the
signal amplitude drops to zero.

The preceding description is seen to correspond to the signals
shown in Fig.\ \ref{fig:Spectra}(a)---the dispersive (in-phase)
and absorptive (quadrature) components and the magnitude of a
Lorentzian with characteristic width $\gamma_t$---as well as to
the lowest order terms of Eqs.\
\eqref{expandedcoscoeff}--\eqref{expandedabscoeff}.

\subsection{Intermediate-field regime}

\begin{figure}
\centering\includegraphics{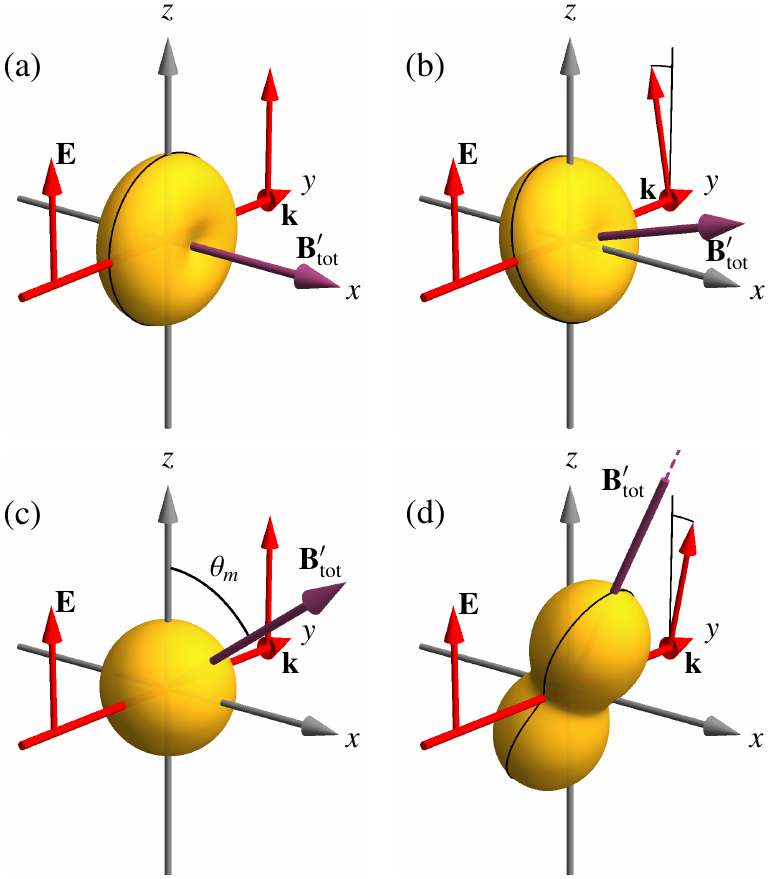}
\caption{Rotating-frame AMPS for intermediate rf
field strength, in the ideal case in which
$\gamma_t\ll\Omega_{\rf}\ll|\Omega_L|$. Optical rotation is
indicated for the instant in which the rotating frame coincides
with the laboratory frame---this means that the generation of
the in-phase component of the signal is shown. (a) When the rf
field is on resonance, the atomic polarization (created along
the $z$-axis) is averaged about the $x$-axis. The resulting
polarization is a ``doughnut'' aligned along the $x$-axis. The
transmission axis, perpendicular to the alignment axis, is
marked with a line on the surface. (b) For small detunings, the
alignment axis follows the effective magnetic field direction.
(c) When the effective magnetic field is at the magic angle
$\theta_m=\arccos(1/\sqrt{3})$ to the light polarization
direction, the atomic polarization is completely averaged out
due to precession. (d) For larger detunings, the polarization
regains its original ``peanut'' shape, and the transmission
axis is along the alignment axis. Media 3 shows the atomic
polarization and in-phase optical rotation as the rf-field
detuning is swept through resonance.} \label{fig:MedPowerAMPS}
\end{figure}

When the rf field is large enough that
$\gamma_t<\Omega_{\rf}<|\Omega_L|$, the rotating-frame
precession frequency is high even at zero detuning. This causes
averaging of the atomic polarization about the magnetic-field
axis. For $\Delta_{\rf}=0$ this is the $x$-axis: polarization
transverse to the $x$-axis is averaged out. However, the
polarization along the $x$-axis is preserved, so that the
$x$-axis becomes the preferred axis for the polarization (Fig.\
\ref{fig:MedPowerAMPS}a). The ``doughnut''-shaped probability
distribution seen in Fig.\ \ref{fig:MedPowerAMPS}(a) is
obtained from the initially pumped ``peanut''-shaped
distribution (Fig.\ \ref{fig:LowPowerOnRes}a) when copies of
the peanut distribution rotated by arbitrary angles about the
$x$-axis are averaged together. Another way to explain the
doughnut shape is to transform to the basis in which the
quantization axis is along $\uv{x}$. In this basis, the
excitation light is $\sigma$ polarized, so that it pumps atoms
out of the bright state consisting of a superposition of the
$m_g=\pm1$ sublevels, and leaves them in the dark state made up
of the opposite superposition, as well as in the $m_g=0$
sublevel. However, due to the precession induced by the
$\uv{x}$ directed magnetic field, atoms oscillate between the
bright and dark superpositions, so that the pump light removes
atoms from the $m_g=\pm1$ sublevels incoherently. The atoms are
then left in the $m_g=0$ sublevel, i.e., the atoms have no
angular-momentum projection on the $x$-axis and are symmetric
about the $x$-axis, as seen in Fig.\ \ref{fig:MedPowerAMPS}(a).

Polarization along the $x$-axis remains in the $xy$ plane as it
precesses around the $z$-axis in the laboratory frame and so
does not induce any optical rotation. However, when the rf
field is tuned slightly away from resonance,
$\V{B}'_{\text{tot}}$ points away from the $x$-axis, and so
also does the averaged atomic polarization (Fig.\
\ref{fig:MedPowerAMPS}b). The polarization then causes optical
rotation. As the polarization in the rotating frame is in the
$xz$ plane, the signal in the laboratory frame is in phase with
the rf field. (The plots in Fig.\ \ref{fig:MedPowerAMPS} are
shown for the instant at which the rotating frame coincides
with the laboratory frame, so that the in-phase component of
the optical rotation is shown.)

To understand the generation of optical rotation in this case,
it is important to note the effect of the different character
of the ensemble polarization. In the low-field case, the
angular-momentum probability distribution has maxima along the
ensemble polarization axis (a peanut), while in the current
case, the distribution has minima (a doughnut). (This result
depends on the particular type of transition considered---for a
$F_g\to F_e=F_g$ transition the situation is reversed.) We can
think of the doughnut shape as an unpolarized distribution
(sphere) with a peanut shape removed. From this viewpoint, it
is reasonable that this ``negative polarization'' produces
rotation of the opposite sign. More concretely, we can note
that, as described above, in a doughnut distribution atoms are
concentrated in the $m_g=0$ sublevel with the quantization axis
along the alignment axis. The state then preferentially absorbs
light that is polarized along the alignment axis. In the
analogy with a polarizing filter, the transmission axis of the
doughnut-shaped probability distribution is transverse to,
rather than along, the alignment axis. This explains the sign
of the rotation shown in Fig.\ \ref{fig:MedPowerAMPS}(b) when
$\V{B}'_{\text{tot}}$ points away from the $x$-axis. (The
transmission axis of each polarization state is marked with a
line on the surfaces plotted in Fig.\ \ref{fig:MedPowerAMPS}.)

\begin{figure}
\centering\includegraphics{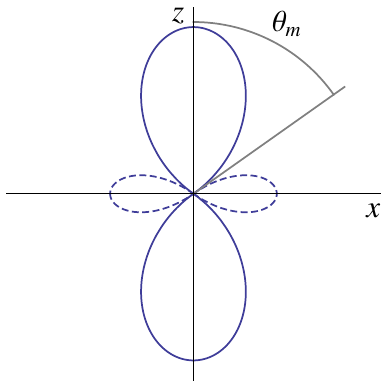}
\caption{Cross section of the AMPS for pure
alignment along the $z$-axis, described by
$Y_{2,0}(\theta,\phi)\propto3\cos^2\theta-1$. Positive function
values are shown as solid lines, negative function values as
dashed lines. The magic angle $\theta_m=\arccos(1/\sqrt{3})$ is
indicated.} \label{fig:MagicAngle}
\end{figure}

As the rf field is tuned farther from resonance,
$\V{B}'_{\text{tot}}$ points farther away from the $x$-axis,
bringing the averaged alignment axis with it. This larger angle
produces a larger optical rotation signal. However, another
trend eventually takes over: because of the shape of the
initially pumped polarization distribution, the amount of
polarization that lies along the magnetic-field direction
decreases. As a result, the amount of averaged polarization is
reduced, tending to reduce the signal. In order to analyze
this, we can plot just the aligned part of the initially pumped
density matrix, neglecting the isotropic part that is included
in Fig.\ \ref{fig:LowPowerOnRes}(a). The surface corresponding
to just the rank $\kappa=2$, $q=0$ polarization moment is
described by the spherical harmonic
$Y_{2,0}(\theta,\phi)\propto3\cos^2\theta-1$, plotted in cross
section in Fig.\ \ref{fig:MagicAngle}.
Negative values of the function are indicated by dashed lines.
There are maximum positive values along $z$ ($\theta=0$), and
maximum negative values in the $xy$ plane ($\theta=\pi/2$). As
$\theta$ moves away from either of these values, the magnitude
of the polarization is reduced. At a particular angle
$\theta_m=\arccos(1/\sqrt{3})$, analogous to the magic angle
observed in nuclear magnetic resonance experiments, the
polarization moment goes to zero. This means that if
$\V{B}'_{\text{tot}}$ is at this angle to the $z$-axis, the
averaged polarization completely cancels, sending the optical
rotation signal to zero (Fig.\ \ref{fig:MedPowerAMPS}c). This
condition corresponds to the additional zero crossings seen in
the in-phase component of Fig.\ \ref{fig:Spectra}(b,c) above
and below the center of the resonance. The direction of the
effective magnetic field in the rotating frame depends on the
rf field strength and the detuning from resonance---as the
field strength is increased, the detuning required to achieve
the magic angle also increases.

When the detuning is even larger, there is once again residual
polarization after averaging about the magnetic-field
direction. However, now the magnetic-field direction is close
enough to the initial alignment axis that the polarization
resulting from averaging resembles the peanut shape of the
initially pumped polarization (Fig.\ \ref{fig:MedPowerAMPS}d).
Thus, the optical-rotation signal in this case is opposite in
sign to that for small detuning and has the same sign as that
for the low-power case.

As the detuning continues to increase, $\V{B}'_{\text{tot}}$
and the averaged atomic alignment point more toward the
$z$-axis, reducing the optical-rotation signal.

The preceding discussion describes the in-phase signal shown in
Fig.\ \ref{fig:Spectra}(b) and (c): a power-broadened
Lorentzian with a narrower feature of the opposite sign in the
center. If the discussion is strictly interpreted, there should
be no quadrature signal in this regime, as the polarization in
the rotating frame is always in the $xz$ plane. Figure
\ref{fig:Spectra} does display (strongly suppressed) quadrature
signals, which are a remnant of the low-field regime.

\subsection{High-field regime}

As $\Omega_{\rf}$ becomes of the same order as $\Omega_L$ or
exceeds it, various higher-order effects appear in the data and
the full theory that cannot be described under the
rotating-wave approximation for the rf field. In particular, a
resonance near $\Omega_L/3$ is seen, as shown in Fig.\
\ref{fig:HarmonicSpectra}. This can be explained as due to ac
Zeeman shifts, which produce evenly spaced sidebands that
result in resonances at odd subharmonics of the lowest-order
resonance. The additional features can also be interpreted as
arising from higher-order resonances between the Larmor
precession and the rf frequency, similar to those seen in
nonlinear magneto-optical rotation with frequency-modulated
light \cite{Alexandrov2002JOSAB}. As the rf field strength
increases, many additional resonances are predicted by the
theory.

\section{Conclusion}

We have conducted a detailed experimental and theoretical
investigation of magneto-optical rotation in the presence of a
strong radio-frequency field. A model has been developed that
allows both the reproduction of the nontrivial line shapes that
appear as the rf field strength is increased, and also the
qualitative understanding of the atomic polarization dynamics
responsible for the signals. The results of this study will be
useful for the analysis of the behavior of atomic magnetometers
exposed to strong radio-frequency fields. The results may also
have applications in the design of efficient methods for the
preparation of specific polarization states of atoms, for
example, atomic states with large alignment but no orientation,
as are used in the work of Ref.\ \cite{DFJacksonKimball2010}.

\section*{Acknowledgments}

We are grateful to M. P. Ledbetter for his help with the
experimental setup. This research was supported by Grant No.\
2006220 from the United States-Israel Binational Science
Foundation (BSF) and by the ONR MURI and NGA NURI programs.
S.G. and E.J.B. were undergraduate exchange students at the
University of California at Berkeley while working on this
project.

\end{document}